\documentclass[12pt]{article}
\usepackage{a4wide}
\usepackage{graphicx}
\def\1{\mbox{l\hspace{-0.53em}1}}
\DeclareMathAlphabet{\mathbbm}{U}{bbm}{m}{n}
  \SetMathAlphabet\mathbbm{bold}{U}{bbm}{bx}{n}

\begin{document}
\def\d{{\rm d}}
\title{\bf Negative parity nonstrange  baryons  in large $N_c$ QCD: quark excitation versus meson-nucleon scattering}
\author{N. Matagne
\thanks{{\it e-mail}: nicolas.matagne@umons.ac.be} ~and Fl.\ Stancu\thanks{{\it e-mail}: fstancu@ulg.ac.be} \\
 {\small $^\ast$ Service de Physique Nucl\'eaire et Subnucl\'eaire, University of Mons,}\\[-6pt]
 {\small Place du Parc, B-7000 Mons, Belgium,} \\ %
 {\small $^\dagger$ Institute of Physics, B5, University of Li\`ege,}\\[-6pt]
 {\small Sart Tilman, B-4000 Li\`ege 1, Belgium}}
\maketitle
\begin{abstract}\noindent
We show that the two complementary pictures of  large $N_c$  baryons - the single-quark orbital excitation 
about a  symmetric core and the 
meson-nucleon resonance -- are compatible for $\ell$ = 3 SU(4) baryons. 
The proof is based on a simple Hamiltonian including operators up to order $\mathcal{O}(N^0_c)$
used previously in the literature for $\ell$ = 1.
\end{abstract}

\section{The status of the $1/N_c$ expansion method}

The large $N_c$ QCD, or alternatively the $1/N_c$ expansion method, proposed by  't Hooft  \cite{HOOFT}
and implemented  by Witten  \cite{WITTEN} became  a valuable tool to study baryon properties 
in terms of the parameter $1/N_c$  where $N_c$ is the number
of colors.  
According to Witten's intuitive picture, a baryon containing $N_c$ quarks 
is seen as a bound state in an average self-consistent potential of a Hartree type 
and the corrections to the Hartree approximation are of order $1/N_c$.

Ten years after 't Hooft's work, Gervais and Sakita  \cite{Gervais:1983wq}
and independently Dashen and Manohar in 1993  \cite{DM} derived a set of consistency conditions for the pion-baryon
coupling constants  which imply that  the large $N_c$ limit of QCD 
has an exact contracted SU(2$N_f$)$_c$ symmetry  
when $N_c \rightarrow \infty $,   $N_f$ being the number 
of flavors.
For ground state baryons the SU(2$N_f$) symmetry is broken by 
corrections proportional to $1/N_c$
\cite{Dashen:1994qi,Jenkins:1998wy}.

Analogous to s-wave baryons, consistency conditions which constrain the strong couplings  
of excited baryons to pions were derived  in Ref. \cite{Pirjol:1997bp}.
These consistency conditions predict the equality between  pion couplings to excited states
and pion couplings to s-wave baryons. These predictions are consistent with the nonrelativistic 
quark model.

A few years later, in the spirit of the Hartree approximation
a procedure for constructing  large $N_c$ baryon wave functions 
with mixed symmetric spin-flavor parts has been proposed
\cite{Goity:1996hk} and an operator analysis was performed for $\ell$ = 1
baryons  \cite{Carlson:1998vx}. 
It was proven that,  for such states,
the SU($2N_f$) breaking occurs at order $N^0_c$, instead of $1/N_c$, as it is the case for  ground and also symmetric 
excited states  $[56, \ell^+]$ (for the latter see Refs. \cite{Goity:2003ab,Matagne:2004pm}).
This procedure has been extended to positive parity nonstrange baryons belonging to the $[70, \ell^+]$ with $\ell$ = 0 and 2
\cite{Matagne:2005gd}. In addition, in Ref.  \cite{Matagne:2005gd},  the dependence of the contribution of the linear term in $N_c$, of the spin-orbit
and of the spin-spin terms in the mass formula was  presented as a function of the excitation energy
or alternatively in terms of the band number $N$.  
Based on this analysis   an impressive global compatibility between the $1/N_c$ expansion and the quark model results  
for $N$ = 0, 1, 2 and 4 \cite{Semay:2007cv} was found
(for a review see Ref. \cite{Buisseret:2008tq}).
More recently the $[70,1^-]$ multiplet was reanalyzed by using an exact wave function, instead of the 
Hartree-type wave function, which allowed to keep control of the Pauli principle at any stage 
of the calculations \cite{Matagne:2006dj}.  The novelty was that the isospin-isospin term, neglected previously
 \cite{Carlson:1998vx} becomes as dominant in $\Delta$ resonances as the spin-spin term in $N^*$ resonances.
  
The purpose of this work is to analyze the compatibility between the $1/N_c$ expansion method in the so-called
${\it quark-shell ~picture}$ and the ${\it resonance~ or~ scattering~ picture}$ defined in the framework of chiral soliton models.
Details can be found in Ref. \cite{Matagne:2011sn}.

\section{Negative parity  baryons}\label{se:excit}

If an excited baryon belongs to a symmetric $[\bf{56}]$-plet
the three-quark system  can be treated similarly to the ground state
in the flavour-spin degrees of freedom, but one has to take into
account the presence of an orbital excitation in the space
part of the wave function  \cite{Goity:2003ab,Matagne:2004pm}.
If the baryon state is described by 
a mixed symmetric representation, $[\bf{70}]$ in SU(6)
notation, the treatment becomes more complicated. 
In particular, the 
resonances up to 2 GeV belong to  $[{\bf 70},1^-]$, $[{\bf 70},0^+]$ or 
$[{\bf 70},2^+]$ multiplets and beyond to 2 GeV to  $[{\bf 70},3^-]$, $[{\bf 70},5^-]$, etc.

In the following we adopt the standard way to study the   $[\bf{70}]$-plets
which, as already mentioned,  is related to the Hartree approximation  \cite{Goity:1996hk}.
An excited baryon is described by a symmetric core plus 
an excited quark coupled to this core, see \emph{e.g.} 
\cite{Carlson:1998vx,Matagne:2005gd,Goity:2002pu,Matagne:2006zf}.
The core is treated in a way similar to that of the ground state.
In this method each SU($2N_f$) $\times$ O(3) generator is  separated
into two parts 
\begin{equation}\label{CORE}
S^i = s^i + S^i_c; ~~~~T^a = t^a + T^a_c; ~~~ G^{ia} = g^{ia} + G^{ia}_c;
~~~ \ell^i = \ell^i_q + \ell^i_c,
\end{equation}
where  $s^i$, $t^a$, $g^{ia}$ and $\ell^i_q$  are the excited 
quark operators and  
$S^i_c$, $T^a_c$, $G^{ia}_c$ and  $\ell^i_c$ the corresponding core operators.

\subsection{The quark-shell picture}

In the quark-shell  picture we use the procedure  of Ref. \cite{COLEB1}, equivalent to that of
Ref. \cite{Pirjol:2003ye}, later extended in Ref. \cite{COLEB2}.
We start from  the leading-order Hamiltonian 
including operators up to order $\mathcal{O}(N^0_c)$ which has the following form 
\begin{equation}\label{TOY}
H = c_1 \ \1  + c_2 \ell \cdot s + c_3 \frac{1}{N_c}\ell^{(2)} \cdot g \cdot G_c
\end{equation}
This operator is defined in the spirit of a Hartree picture (mean field) 
where the matrix elements of the first term are proportional to $ N_c$ on all baryons \cite{WITTEN}. 
The spin-orbit term $\ell \cdot s$ 
which is a one-body operator and the  third term - a two-body operator containing the tensor  
$\ell^{(2)ij}$ of O(3) - have matrix elements of order $\mathcal{O}(N^0_c)$. The neglect of $1/N_c$
corrections  in the $1/N_c$ expansion makes sense for the comparison with the scattering picture
in the large $N_c$ limit, described in the following section.

One can see that the  Hamiltonian  (\ref{TOY}) reproduces  
the characteristic $N_c$ scaling for
the excitation energy of baryons which is $N^0_c$   \cite{WITTEN}.

\subsubsection{The nucleon case}

In large $N_c$ the color part of the wave function is antisymmetric so that the
orbital-spin-flavor part must be symmetric to satisfy the Pauli principle.  A quanta of orbital excitation requires 
the orbital part to be mixed  symmetric, the lowest state having the partition $[N_c-1,1]$.
We have the following $[N_c-1,1]$ spin-flavor  ($SF$) states which form a symmetric state 
with the orbital $\ell$ = 3 state of partition $[N_c - 1,1]$  

\begin{enumerate}
\item
$\left[N_c - 1, 1\right]_{SF} = \left[\frac{N_c+1}{2}, \frac{N_c - 1}{2}\right]_{S} \times  \left[\frac{N_c+1}{2}, \frac{N_c - 1}{2}\right]_{F} $, $N_c  \geq 3$ \\
with $S = 1/2$ and $J = 5/2, 7/2$
\item
$\left[N_c - 1, 1\right]_{SF} = \left[\frac{N_c+3}{2}, \frac{N_c - 3}{2}\right]_{S} \times  \left[\frac{N_c+1}{2}, \frac{N_c - 1}{2}\right]_{F} $,   $N_c  \geq 3$ \\  
with $S = 3/2$ and $J = 3/2, 5/2, 7/2, 9/2$.
\end{enumerate} 
They give rise to matrices of a given $J$ either $2 \times 2$ or $1 \times 1$ depending on the 
multiplicity of  $J$. States of symmetry $[N_c - 1, 1]_{SF}$ with  
  $S = 5/2$,  like for $\Delta$ (see below), which together with $\ell = 3$ could give rise  to $J = 11/2$, 
   are not allowed for $N$, by inner products of the permutation group
  \cite{Stancu:1991rc}. Therefore the experimentally observed resonance $N(2600) I_{11/2}$ should belong to the $N = 5$ band ($\ell$ = 5). 
For $N_c$ = 3 the above states correspond to the $^28$ and $^48$ multiplets of SU(2) $\times$ SU(3) respectively.

\subsubsection{The $\Delta$ case}

In this case the Pauli principle allows the following states
\begin{enumerate}
\item
$\left[N_c - 1, 1\right]_{SF} = \left[\frac{N_c+1}{2}, \frac{N_c - 1}{2}\right]_{S} \times  \left[\frac{N_c+3}{2}, \frac{N_c - 3}{2}\right]_{F} $,   $N_c  \geq 3$ \\  
with $S = 1/2$ and $J = 5/2, 7/2$,
\item
$\left[N_c - 1, 1\right]_{SF} = \left[\frac{N_c+3}{2}, \frac{N_c - 3}{2}\right]_{S} \times \left[\frac{N_c+3}{2}, \frac{N_c - 3}{2}\right]_{F} $,   $N_c  \geq 5$ \\  
with $S = 3/2$ and $J = 3/2, 5/2, 7/2, 9/2$,
\item
$\left[N_c - 1, 1\right]_{SF} = \left[\frac{N_c+5}{2}, \frac{N_c - 5}{2}\right]_{S} \times \left[\frac{N_c+3}{2}, \frac{N_c - 3}{2}\right]_{F} $,   $N_c  \geq 7$ \\ 
with $S = 5/2$ and $J = 1/2, 3/2, 5/2, 7/2, 9/2, 11/2$.
\end{enumerate} 
As above, they indicate the size of a matrix of fixed $J$ for the Hamiltonian (\ref{TOY}).  For example, 
the matrix of $\Delta_{5/2}$ is 3$\times$3, because all three
states can have $J = 5/2$. 
For $N_c = 3$ the first state belongs to the $^210$ multiplet.
The other two types of states do not appear in the real world with $N_c = 3$.
Note that both for $N_J$ and $\Delta_J$ states the size of a given matrix equals the multiplicity of the corresponding
state indicated in Table 1 of Ref. \cite{COLEB2} for $\ell = 3$. 

The Hamiltonian (\ref{TOY}) is diagonalized in the bases defined above. Let us denote the eigenvalues either by
$m^{(i)}_{N_J}$ or $m^{(i)}_{\Delta_J}$ with $i$ = 1, 2 or 3, depending on how many eigenvalues are at a fixed $J$.
The Hamiltonian has analytical solutions, all eigenvalues being linear functions in the coefficients $c_1$, $c_2$ and $c_3$.
It is remarkable that the 18 available eigenstates with $\ell$ = 3 fall into three degenerate multiplets,
like for $\ell$ = 1.  If the degenerate masses are denoted by $m'_2$,  $m_3$ and $m_4$ 
we have
\begin{equation}\label{mass3}
m'_2 =   m^{(1)}_{\Delta_{1/2}}  =  m^{(1)}_{N_{3/2}} = m^{(1)}_{\Delta_{3/2}} =  m^{(1)}_{N_{5/2}}  = m^{(1)}_{\Delta_{5/2}} = m^{(1)}_{\Delta_{7/2}} ,
\end{equation}
\begin{equation}\label{mass4}
m_3 = m^{(2)}_{\Delta_{3/2}} =  m^{(2)}_{N_{5/2}} =  m^{(2)}_{\Delta_{5/2}} =  m^{(1)}_{N_{7/2}} =  m^{(2)}_{\Delta_{7/2}} = m^{(1)}_{\Delta_{9/2}},
\end{equation}
\begin{equation}\label{mass5}
m_4 = m^{(3)}_{\Delta_{5/2}} =  m^{(2)}_{N_{7/2}} = m^{(3)}_{\Delta_{7/2}} =  m^{(1)}_{N_{9/2}}  = m^{(2)}_{\Delta_{9/2}}
= m^{(1)}_{\Delta_{11/2}},
\end{equation}
where
\begin{equation}  
m'_2 = c_1 N_c - 2 c_2 - \frac{3}{4} c_3, 
\end{equation}  
\begin{equation}  
m_3 = c_1 N_c - \frac{1}{2} c_2 + \frac{15}{16} c_3, 
\end{equation}
\begin{equation}  
m_4 = c_1 N_c + \frac{3}{2} c_2 - \frac{5}{16} c_3. 
\end{equation}
The notation $m'_2$ is used to distinguish this eigenvalue from $m_2$ of Ref. \cite{COLEB1}.

In the following subsection we shall see that the scattering picture gives an identical pattern
of degeneracy in the quantum numbers, but the resonance mass is not quantitatively defined. 
Therefore only a qualitative compatibility can be established.

\subsection{The meson-nucleon scattering picture}

Here we are concerned with nonstrange baryons, as above, and look for a degeneracy pattern in the resonance picture.
The starting point in this analysis   are the linear relations 
of the S matrices  $S^{\pi}_{LL'RR'IJ}$ and  $S^{\eta}_{LRJ}$ of $\pi$ and $\eta$ scattering off 
a ground state baryon in terms of $K$-amplitudes. They are given by the following equations  \cite{COLEB1,COLEB2}
\begin{equation}\label{pi}
S^{\pi}_{LL'RR'IJ} = \sum_K ( - 1)^{R'-R} \sqrt{(2R+1)(2R'+1)} (2K+1)
\left\{\begin{array}{ccc}
        K& I & J \\
	R' & L' & 1
      \end{array}\right\} 
 \left\{\begin{array}{ccc}
        K& I & J \\
	R & L & 1
      \end{array}\right\}  
      s^{\pi}_{KLL'},    
\end{equation}
and
\begin{equation}\label{eta}
S^{\eta}_{LRJ} = \sum_K \delta_{KL}\delta(LRJ) s^{\eta}_{K},
\end{equation}
where  $s^{\pi}_{KL'L}$ and $s^{\eta}_{K}$ are the reduced amplitudes. 
The notation is as follows. For $\pi$ scattering $R$ and $R'$ are the spin of the incoming and outgoing baryons 
respectively ($R$ =1/2 for $N$ and $R$ = 3/2 for $\Delta$), $L$ and $L'$ are the partial wave angular momentum of the
incident and final $\pi$ respectively (the orbital angular momentum $L$ of $\eta$ remains unchanged), 
$I$ and $J$ represent the total isospin and total angular momentum
associated to a given resonance 
and $K$ is the 
magnitude  of the ${\it grand}$ ${\it spin}$ $\vec{K} = \vec{I} + \vec{J}$.
The $6j$ coefficients imply four triangle rules $\delta(LRJ)$, $\delta(R1I)$, $\delta(L1K)$ and 
$\delta(IJK)$.

These equations were first derived in the context  of the chiral soliton model 
\cite{HAYASHI,MAPE}
where 
the mean-field breaks the rotational and isospin symmetries, so that $J$ and $I$ are not
conserved but the ${\it grand}$ ${\it spin}$  $K$ is conserved and excitations can be labelled by $K$.
These relations are exact in large $N_c$ QCD and are independent of any model assumption.

The meaning of Eq. (\ref{pi}) is that there are more amplitudes $S^{\pi}_{LL'RR'IJ}$ than there are $s^{\pi}_{KLL'}$
amplitudes.  The reason is that the $I J$ as well as the $R R'$ dependence is contained only  in the geometrical
factor containing the two $6j$ coefficients.  
Then, for example, in the $\pi N$ scattering, in order for a resonance to occur in one channel there 
must be a resonance in at least
one of the contributing amplitudes $s^{\pi}_{KLL'}$. But as $s^{\pi}_{KLL'}$ contributes
in more than one channel,  all these channels resonate at the same energy and this implies degeneracy
in the excited spectrum.  From the chiral soliton model there is no reason to suspect degeneracy 
between different $K$ sectors.

From the meson-baryon scattering relations  (\ref{pi}) and   (\ref{eta})
three sets of degenerate states have been found for $\ell$ = 1 orbital 
excitations \cite{COLEB1}.
There is a clear correspondence
between these sets and 
 the three towers of states \cite{COLEB1,Pirjol:2003ye}
 of the excited quark picture provided by the 
symmetric core + excited quark scheme \cite{Carlson:1998vx}.
They correspond to $K = 0, 1$ and 2 in the resonance picture.
But the resonance picture also provides a $K = 3$ due to the amplitude 
$s^{\pi}_{322}$. 
As this is different from the other $s^{\pi}_{KL'L}$ , in Ref. \cite{COLEB1}
it was interpreted as belonging to the $N = 3$ band. 

Here we extend the work of Ref. \cite{COLEB1,COLEB2} to $\ell = 3$ excited states which 
belong to the $N = 3$ band. 
The partial wave amplitudes of interest and their expansion  
in terms of $K$-amplitudes from Eqs.~(\ref{pi}) and (\ref{eta}) can be found in Tables I-III of
Ref.  \cite{Matagne:2011sn}.  They correspond   
to   $L = L' = 2$,   $L = L' = 4$ and $L = L' = 6$ respectively. 
From those tables one can infer the following degenerate towers of states
with their contributing amplitudes 
\begin{eqnarray}
\Delta_{1/2}, \; \; \;   N_{3/2} , \; \; \;   \Delta_{3/2} , \; \; \;   N_{5/2} , \; \; \;  \Delta_{5/2} , \; \; \;  \Delta_{7/2} , \; \; \;
&~&
 (s_{222}^\pi, s_{2}^\eta) , \label{s2p}\\
 \Delta_{3/2} , \; \; \;   N_{5/2} , \;  \; \;  \Delta_{5/2} , \; \;  \; N_{7/2} , \; \; \;
\Delta_{7/2} , \; \; \;  \Delta_{9/2} , 
 &~& (s_{3 2 2}^\pi,  s_{3 4 4}^\pi) , \label{s1}\\
 \Delta_{5/2} , \;  \; \;  N_{7/2} , \;  \; \; \Delta_{7/2} , \; \; \; N_{9/2} , \; \; \;  \Delta_{9/2} ,
 \;  \; \; \Delta_{11/2} , 
 &~& ( s_{4 4 4}^\pi, s_{4}^\eta ) , \label{s2} \\
\Delta_{7/2} , \; \; \; N_{9/2} , \; \;  \; \Delta_{9/2} ,
 \; \; \; \Delta_{11/2} , \; \; 
&~&
(s^\pi_{5 4 4 },  s^\pi_{5 6 6 }), \;  \;  \label{s3} \\
\Delta_{9/2}, \; \; \; 
\Delta_{11/2} , 
&~& (s^\pi_{6 6 6 }, s_{6}^\eta ) \label{s4}
\end{eqnarray}
associated to  $K = 2, 3, 4, 5$ and 6 respectively.

We can  compare the towers (\ref{s2p})-(\ref{s4})
 with the quark-shell model results of (\ref{mass3})-(\ref{mass5}).
The first observation is that  the agreement of 
 (\ref{s2p}) ($K = 2$) with (\ref{mass3}), of
 (\ref{s1}) ($K = 3$) with  (\ref{mass4})  and  of (\ref{s2})  ($K = 4$)  with (\ref{mass5}) 
is perfect regarding the quantum numbers.
Second, we note that the resonance picture can have poles with $K = 5, 6$ 
 which infer the towers (\ref{s3}) and (\ref{s4}).  They have no counterpart
in the quark-shell picture for $\ell = 3$. 
But there is no problem because the poles with $K = 5, 6$  can belong to a higher band, 
namely $N = 5$ ($\ell = 5$) without spoiling the compatibility. 

Comparing these results with those of Ref. \cite{COLEB2} one can conclude that 
one can associate a common $K = 2$ to $\ell = 1$ and $\ell = 3$.  For this value of $K$ 
the triangular rule $\delta( K \ell 1)$ proposed in Ref \cite{COLEB2} is satisfied.
The quark-shell picture brings however more information than the resonance picture
due to the fact that it implies  an energy dependence via the $\ell$ dependence which
measures the orbital excitation.  Note that $m'_2$ is different from $m_2$ of $\ell = 1$
\cite{COLEB1,Pirjol:2003ye}. Because in the resonance 
picture they stem from the same amplitude $s^{\pi}_{222}$, one expects that this
amplitude possesses two poles at two distinct energies, in order to have compatibility.
Thus the number of poles of the reduced amplitudes $s^{\pi}_{KLL}$ remains an open question.

We anticipate that a similar situation will appear  for every value of $K$ 
associated to two distinct values of $\ell$, satisfying the $\delta(K \ell 1)$ rule, for example, for
$K$ = 4 which is  common to $\ell$ = 3  and $\ell$ = 5.


\section{Conclusions}\label{se:concl}
We have compared two alternative pictures for baryon resonances consistent with 
large the $N_c$ QCD limit and found that the two pictures are compatible for $\ell$ = 3 
excited states, as it was the case for $\ell$ = 1. The quark-shell picture is practical and successful 
in describing known resonances 
and in predicting other members of the excited octets and decuplets. But the extended symmetry 
SU(2$N_f$) $\times$ O(3) where O(3),  which is essential to include orbital excitations, does not have
a direct link to large  $N_c$. 
On the other hand the scattering picture is close to experimental analysis but it is not clear where
the pole positions should lie. It is however very encouraging that the two pictures give 
sets of  degenerate states with identical quantum numbers when one works at order 
$\mathcal{O}(N^0_c)$. It is a qualitative proof that the spin-flavor picture is valid
and useful for baryon phenomenology. 

 
\end{document}